\documentclass[a4paper,12pt]{article}

\usepackage{amssymb,amsmath,amsthm}
\usepackage[totalwidth=17cm,totalheight=24cm]{geometry}
\usepackage{graphicx}

\newcommand{\C}{{\mathbb C}}
\newcommand{\R}{{\mathbb R}}
\newcommand{\Z}{{\mathbb Z}}
\newcommand{\Hq}{{\mathbb H}}
\newcommand{\Oc}{{\mathbb O}}
\newcommand{\X}{{\mathbb X}}
\newcommand{\Y}{{\mathbb Y}}

\newcommand{\id}{{\mathbb I}}
\newcommand{\im}{{\rm i\,}}

\newcommand{\lla}{\langle\!\langle}
\newcommand{\rra}{\rangle\!\rangle}

 \theoremstyle{plain}
  \newtheorem{theorem}{Theorem}

  \theoremstyle{definition}

  \theoremstyle{remark}
  \newtheorem{remark}{Remark}

\newcommand{\be}{\begin{eqnarray}}
\newcommand{\ee}{\end{eqnarray}}

\begin{document}
 \pagestyle{plain}
\title{${\rm Spin}(11,3)$, particles and octonions}

\author{Kirill Krasnov\\ {}\\
{\it School of Mathematical Sciences, University of Nottingham, NG7 2RD, UK}}

\date{April 2021}
\maketitle

\begin{abstract}\noindent The fermionic fields of one generation of the Standard Model, including the Lorentz spinor degrees of freedom, can be identified with components of a single real 64-dimensional semi-spinor representation $S_+$ of the group ${\rm Spin}(11,3)$. We describe an octonionic model for ${\rm Spin}(11,3)$ in which the semi-spinor representation gets identified with $S_+=\Oc\otimes\tilde{\Oc}$, where $\Oc,\tilde{\Oc}$ are the usual and split octonions respectively. It is then well-known that choosing a unit imaginary octonion $u\in {\rm Im}(\Oc)$ equips $\Oc$ with a complex structure $J$. Similarly, choosing a unit imaginary split octonion $\tilde{u}\in{\rm Im}(\tilde{\Oc})$ equips $\tilde{\Oc}$ with a complex structure $\tilde{J}$, except that there are now two inequivalent complex structures, one parametrised by a choice of a timelike and the other of a spacelike unit $\tilde{u}$. In either case, the identification $S_+=\Oc\otimes\tilde{\Oc}$ implies that there are two natural commuting complex structures $J, \tilde{J}$ on $S_+$. Our main new observation is that the subgroup of ${\rm Spin}(11,3)$ that commutes with both $J, \tilde{J}$ on $S_+$ is the direct product ${\rm Spin}(6)\times{\rm Spin}(4)\times{\rm Spin}(1,3)$ of the Pati-Salam and Lorentz groups, when $\tilde{u}$ is chosen to be timelike. The splitting of $S_+$ into eigenspaces of $J$ corresponds to splitting into particles and anti-particles. The splitting of $S_+$ into eigenspaces of $\tilde{J}$ corresponds to splitting of Lorentz Dirac spinors into two different chiralities. We also study the simplest possible symmetry breaking scenario with the "Higgs" field taking values in the representation that corresponds to 3-forms in $\R^{11,3}$. We show that this Higgs can be designed to transform as the bi-doublet of the left/right symmetric extension of the SM, and thus breaks ${\rm Spin}(11,3)$ down to the product of the SM, Lorentz and ${\rm U}(1)_{B-L}$ groups, with the last one remaining unbroken. This 3-form Higgs field also produces the Dirac mass terms for all the particles. 
\end{abstract}

\section{Introduction}

The suggestion that octonions may provide a useful language for describing elementary particles is almost as old as the Standard Model (SM) itself, see \cite{Gunaydin:1973rs} and references therein.\footnote{Note, however, that in this reference, somewhat confusingly, the usual octonion division algebra written in a complex basis is referred to as the split octonion algebra.} Thus, it was observed early on that the 8-dimensional space of octonions, after a complex structure is chosen, splits as $\Oc=\C\oplus\C^3$, with the first factor naturally corresponding to leptons and the second one describing quarks. The subgroup of the group of automorphisms of the octonions $G_2$ preserving this split is precisely the strong gauge group ${\rm SU}(3)$. 

Many authors have since been fascinated by a possible link between elementary particles and octonions. The literature on this topic has grown to be large, but also hard to summarise because the subject has never became a communal effort. Individuals that did work on it had different backgrounds and used their own distinct language, making a review of what has been achieved difficult. There is currently a renewed interest in these topics, as is manifested in particular by  \cite{Workshop}. Some of the most notable past developments in relation to this topic are works by Dixon e.g. \cite{Dixon-book}, \cite{Dixon:1999rg}, \cite{Dixon:2010px}, as well as works by Dray and collaborators, see e.g. \cite{Manogue:2009gf} and references therein. More recent works are those by Furey e.g. \cite{Furey:2016ovx}, \cite{Furey:2018yyy} and by Dubois-Violette and Todorov \cite{Dubois-Violette:2016kzx}, \cite{Todorov:2018mwd}, \cite{Dubois-Violette:2018wgs}, \cite{Todorov:2018yvi}, \cite{Todorov:2019hlc}. Octonions play a prominent role in all these papers. 

While some of the papers on this subject only attempt to describe the "internal" particle degrees of freedom on which the gauge group of the SM acts, others are trying to put the Lorentz spinor degrees of freedom into the picture as well.  The basic observation in this regard is that spinors of (psuedo-) orthogonal groups ${\rm Spin}(2n)$, when restricted to their subgroup ${\rm Spin}(2k)\times {\rm Spin}(2l), k+l=n$ decompose as spinors with respect to both ${\rm Spin}(2k), {\rm Spin}(2l)$ factors. This, in particular, means that a spinor of a (pseudo-) orthogonal group in 14 dimensions ${\rm Spin}(14)$, when restricted to ${\rm Spin}(10)\times {\rm Spin}(4)$ will behave as a spinor with respect to both of these groups. As far as we are aware, the first mention of the possibility of such a unification of Lorentz spinor and "internal" (isospin and colour) degrees of freedom is in \cite{Percacci:1984ai}. 

The number of real functions that is needed to describe one generation of the SM fermions is 64. This is the dimension of the Weyl, i.e. semi-spinor representation in 14 dimensions. However, for most possible signature combinations, the arising representation is complex 64-dimensional. The only two cases that admit real, i.e. Majorana semi-spinors in this dimension are the split case ${\rm Spin}(7,7)$ and ${\rm Spin}(11,3)$. The case ${\rm Spin}(11,3)$ was studied in works \cite{Nesti:2009kk}, \cite{Lisi:2010uw}. The split case ${\rm Spin}(7,7)$ was advocated in an earlier work of this author \cite{Krasnov:2018txw}.  

The purpose of this paper is to revisit the idea that all field components of one generation of the SM particles can be described using a real Weyl spinor of a pseudo-orthogonal group in 14 dimensions. The new twist that we add is an observation that there is a natural octonionic model for both ${\rm Spin}(7,7)$ and ${\rm Spin}(11,3)$. One can then consider various different complex structures on $S_+$ arising from the octonions, similar to what was done in \cite{Krasnov:2019auj} in the context of ${\rm Spin}(9)$. We will see that there are two natural commuting complex structures, both corresponding to a choice of a unit imaginary octonion, and their joint commutant in ${\rm Spin}(11,3)$ is the direct product of the Pati-Salam and Lorentz groups. There is also a version of the construction that gives the product of the gauge group of the left/right-symmetric extension of the SM times the Lorentz group. It is the simplicity of this characterisation of the groups so relevant for physics that prompted us to write this paper. 

There are many different ways to describe ${\rm Spin}(7,7), {\rm Spin}(11,3)$. There is always the Killing-Cartan description with roots and weights. Unfortunately, it does become difficult to work with for groups of such high dimension. For orthogonal groups, one can instead proceed starting with the corresponding Clifford algebras. One standard way to obtain these is by taking tensor products of $\gamma$-matrices in lower dimensions. This is indeed the approach that was followed in \cite{Lisi:2010uw}, but it also becomes quite cumbersome. It is more efficient to restrict to some carefully chosen commuting subgroups and build a model for the big group based on the representation theory of the smaller groups. Thus, the semi-spinor representations of ${\rm Spin}(8)$ can be identified with octonions $\Oc$, similarly for the semi-spinor representations of the split ${\rm Spin}(4,4)$ that can be identified with $\tilde{\Oc}$. One can then restrict ${\rm Spin}(11,3)$ to ${\rm Spin}(8)\times{\rm Spin}(3,3)$, and similarly ${\rm Spin}(7,7)$ to ${\rm Spin}(4,4)\times{\rm Spin}(3,3)$, and then view the semi-spinor representation of ${\rm Spin}(11,3)$ and ${\rm Spin}(7,7)$ as the tensor product of spinors of ${\rm Spin}(8), {\rm Spin}(3,3)$ and ${\rm Spin}(4,4), {\rm Spin}(3,3)$ respectively. The semi-spinor representation of ${\rm Spin}(3,3)$ is a copy of $\R^4$. This means that the semi-spinor representation of ${\rm Spin}(11,3)$ can be viewed as 8 copies of the octonions, and that of ${\rm Spin}(7,7)$ can be viewed as 8 copies of the split octonions. This formalism is useful and can be developed further, but there is an even more efficient description.

What seems to be the most natural description of ${\rm Spin}(7,7), {\rm Spin}(11,3)$ arises by restricting to the groups ${\rm Spin}(7)\times{\rm Spin}(7)\subset {\rm Spin}(7,7)$ and ${\rm Spin}(7)\times{\rm Spin}(4,3)\subset {\rm Spin}(11,3)$.\footnote{In the case of  ${\rm Spin}(7,7)$ there is also the possibility ${\rm Spin}(3,4)\times{\rm Spin}(4,3)\subset {\rm Spin}(7,7)$.} The spinor representation of ${\rm Spin}(7)$ can be identified with octonions $\Oc$, and the spinor representation of ${\rm Spin}(4,3)$ with split octonions $\tilde{\Oc}$. This means that a semi-spinor of ${\rm Spin}(7,7)$ can be identified with $\Oc\otimes\Oc$ (or with $\tilde{\Oc}\otimes\tilde{\Oc}$), and a semi-spinor of ${\rm Spin}(11,3)$ can be identified with $\Oc\otimes\tilde{\Oc}$. It is this description and the corresponding models for ${\rm Spin}(7,7), {\rm Spin}(11,3)$ that we will develop in the main text. 

The resulting model for ${\rm Spin}(7,7), {\rm Spin}(11,3)$ can be phrased more generally, as it describes some other pseudo-orthogonal groups as well. It can be summarised in the following proposition:
\begin{theorem} Let ${\mathbb X},{\mathbb Y}$ be a pair of division algebras from the set $\C,\tilde{\C},\Hq,\tilde{\Hq},\Oc,\tilde{\Oc}$, where $\C,\tilde{\C}$ are the usual and split complex numbers, and $\Hq$ and $\tilde{\Hq}$ are the usual and split quaternions respectively. The matrices taking values in ${\rm End}(({\mathbb X}\otimes{\mathbb Y})^2)$ 
\be\label{Gammas}
\Gamma_x = \left( \begin{array}{cc} 0 & L_x\otimes \id \\ L_{\bar{x}}\otimes \id & 0 \end{array}\right), \quad x\in{\rm Im}({\mathbb X}), \qquad
\Gamma_y = \left( \begin{array}{cc} 0 & -\id\otimes L_y \\ \id\otimes L_{\bar{y}} & 0 \end{array}\right), \quad y\in{\rm Im}({\mathbb Y})
\ee
where $L_x, L_y$ are the operators of left multiplication by $x\in{\rm Im}({\mathbb X}), y\in{\rm Im}({\mathbb Y})$, and bar denotes conjugation in ${\mathbb X},{\mathbb Y}$, satisfy 
\be\label{cliff-XY-rels}
\Gamma_x^2=|x|^2 \id, \quad \Gamma_y^2= -|y|^2 \id, \\ \nonumber
\Gamma_x\Gamma_y + \Gamma_y \Gamma_x=0.
\ee 
They thus generate the Clifford algebra of signature that can be symbolically expressed as
\be\label{signature}
| {\rm Im}({\mathbb X})|^2 - | {\rm Im}({\mathbb Y})|^2.
\ee
\end{theorem}
A proof is by direct verification.
\begin{remark} The construction in Theorem 1 is very closely related to that of $2\times 2$ matrix algebras in \cite{Barton:2000ki}. Our construction is a subcase of the more general construction in this reference in the sense that larger groups in \cite{Barton:2000ki} can be obtained by adding to the set of Clifford generators of Theorem 1 two more generators. This adds one positive definite and one negative definite direction. Results in \cite{Barton:2000ki}, see also \cite{Dray:2014mba}, are then reproduced. We will return to this in the main text.
\end{remark}
\begin{remark} This construction of the Clifford algebra identifies spinors with $({\mathbb X}\otimes{\mathbb Y})^2$, and thus a single copy of ${\mathbb X}\otimes{\mathbb Y}$ with semi-spinors.
This immediately gives a description of the Lie algebra in terms of matrices in ${\rm End}({\mathbb X}\otimes{\mathbb Y})$. This will be described in the main text, for the cases of interest.
\end{remark}
\begin{remark} The case of ${\rm Cliff}_{11,3}$ corresponds to ${\mathbb X}=\Oc, {\mathbb Y}=\tilde{\Oc}$. The split case ${\rm Cliff}_{7,7}$ can be described by both ${\mathbb X}=\Oc, {\mathbb Y}=\Oc$, which corresponds to restricting to the ${\rm Spin}(7)\times{\rm Spin}(7)$ subgroup, as well as by ${\mathbb X}=\tilde{\Oc}, {\mathbb Y}=\tilde{\Oc}$, which corresponds to restricting to ${\rm Spin}(3,4)\times{\rm Spin}(4,3)$. Both are useful descriptions and capture two different types of spinor orbits in the $(7,7)$ case. 
\end{remark}
\begin{remark}
Note that there is no way to get the compact case ${\rm Spin}(14)$ by this construction, at least not by keeping the construction real. Indeed, the semi-spinor representations of this group are complex 64-dimensional. One would need to introduce factors of the imaginary unit in one of the two $\Gamma$-matrices to counteract the unavoidable minus sign in (\ref{signature}). Allowing for such factors of imaginary unit generalises the construction of Theorem 1 to an even larger set of groups. 
\end{remark}
Further implications of the construction of Theorem 1 will be discussed in the main text.

With this result at hand we can identify the semi-spinor representation of ${\rm Spin}(11,3)$, and thus components of one fermion generation of the SM, with $\Oc\otimes\tilde{\Oc}$. The Lie algebra of ${\rm Spin}(11,3)$ and thus of its various subgroups of interest can be obtained by taking the commutators of the $\Gamma$-matrices (\ref{Gammas}). This gives an explicit and useful description, as we will further illustrate in the main text. 

This gives an ultimate "kinematic" unification of all known to us components of fermions of one generation in such an elegant structure as $\Oc\otimes\tilde{\Oc}$. But it does not answer the crucial question of any such unification approach: What breaks the big symmetry group such as ${\rm Spin}(11,3)$ to the product of groups we know act on elementary particles in Nature? The important part of this question is what singles out the Lorentz group ${\rm Spin}(1,3)$ that we know plays a very different role from the gauge groups appearing in the Standard Model. Indeed, the group ${\rm Spin}(11,3)$ we have considered mixes the SM gauge group with the transformations from the Lorentz group. We don't know of any such mixed transformations playing any role in our physical theories. So, any "unification" scheme that puts together the SM gauge group with the Lorentz group but does not explain how these two factors appear is largely useless. 

We do not have a satisfactory answer to these questions, but there are some interesting observations that can be made. Let us first recall the more more conventional GUT scenarios.

In ${\rm Spin}(10)$ unification, all fermionic components of one generation (but not including the Lorentz spinor components) can be elegantly described by a single complex 16-dimensional representation. The question is then what breaks the ${\rm Spin}(10)$ symmetry down to the ones we see in the SM. In all considered models, see e.g. \cite{DiLuzio:2011mda} for a nice recent review, the symmetry breaking happens in two stages. The first stage is to break ${\rm Spin}(10)$ down to either the Pati-Salam group ${\rm Spin}(6)\times{\rm Spin}(4)$, or down to the Georgi-Glashow group ${\rm SU}(5)$. Only after this one breaks the remaining symmetry down to the SM gauge group, which is then further broken to ${\rm SU}(3)\times{\rm U}(1)$, the unbroken gauge group of the SM. All these breakings are carried out using various Higgs fields, and no Higgs field taking values in a single irreducible representation of ${\rm Spin}(10)$ is sufficient for the purpose. 

It is known that the breaking of ${\rm Spin}(10)$ to either the Pati-Salam group ${\rm Spin}(6)\times{\rm Spin}(4)$ or to ${\rm SU}(5)\times{\rm U}(1)={\rm U}(5)$ can be carried out using the Higgs in the ${\bf 210}$ representation, see e.g. the diagram in Section 1.4.5 in \cite{DiLuzio:2011mda}. Alternatively, this stage of the symmetry breaking can be reformulated by saying that there exists complex structures in the space of ${\rm Spin}(10)$ semi-spinors whose commutants are either the Pati-Salam or ${\rm U}(5)$. Both of these complex structures are closely related to octonions, and can be reformulated using the octonionic language that exists for ${\rm Spin}(10)$, see e.g. \cite{Bryant} for a description of the octonionic model of ${\rm Spin}(10)$. What matters for our discussion here is that the phenomenologically interesting complex structures that break either to Pati-Salam or to ${\rm U}(5)$ can be both parametrised by a choice of a unit imaginary octonion. 

For the case of the complex structure that breaks ${\rm Spin}(10)$ down to the Pati-Salam group this can be explained without any further background. The octonionic model of ${\rm Spin}(10)$ proceeds by identifying 8 of the 10 directions of the Clifford algebra ${\rm Cliff}_{10}$ with octonions, thus leaving out two of the directions as special. There is another direction that is singled out as special, which corresponds to the identity element in $\Oc$. A choice of a unit imaginary octonion $u\in{\rm Im}(\Oc)$ gives yet another special direction. From this data one can easily construct a complex structure on semi-spinors whose commutant is the Pati-Salam group ${\rm Spin}(6)\times{\rm Spin}(4)$. Indeed, the product of any two $\gamma$-matrices in ${\rm Cliff}_{10}$ gives an operator that squares to minus the identity and preserves the space of semi-spinors. However, the commutant of such a complex structure is ${\rm Spin}(8)\times{\rm Spin}(2)$, so this is not what we want. The next in complexity operator that preserves the space of semi-spinors is given by the product of four of the $\gamma$-matrices. However, this squares to plus the identity, and so is not a complex structure. However, the next case is the product of six different $\gamma$'s. This squares to minus the identity, and so gives a complex structure. It breaks ${\rm Spin}(10)$ down to ${\rm Spin}(6)\times{\rm Spin}(4)$. Finally, it can clearly be parametrised by a choice of a unit imaginary octonion $u$. Indeed, as we already discussed, together with $u$ we have four special directions in ${\rm Cliff}_{10}$, and thus also remaining 6 directions. These are the directions in $\Oc$ that are not in the space ${\rm Span}(\id,u)$. The product of the corresponding $\gamma$-matrices is the desired complex structure that breaks ${\rm Spin}(10)$ to the Pati-Salam group. 

Our main new observation in this paper is that there exist very analogous complex structures on ${\rm Spin}(11,3)$. These complex structures can also be parametrised by a choice of unit imaginary octonions, but now in both $\Oc,\tilde{\Oc}$. A choice of $u\in {\rm Im}(\Oc)$ still singles out 6 of the $\gamma$-matrices in ${\rm Cliff}_{11,3}$, corresponding to the octonions not in ${\rm Span}(\id,u)\subset \Oc$. This gives the complex structure $J$. Its commutant in ${\rm Spin}(11,3)$ is ${\rm Spin}(6)\times{\rm Spin}(5,3)$, which is not interesting in itself. 

Let us now consider the possibilities that arise because we have the $\tilde{\Oc}$ factor in our octonionic model for ${\rm Spin}(11,3)$. The relevant Clifford algebra is ${\rm Cliff}_{4,3}$, where the first 4 directions can be referred to as timelike. These are the negative-definite directions of the corresponding quadratic form. Now a choice of a unit imaginary octonion $\tilde{u}\in{\rm Im}(\tilde{\Oc})$ is a choice of either a timelike or a spacelike direction. Choosing $\tilde{u}$ to be spacelike gives us the complex structure that is given by the product of 6 $\gamma$-matrices not in ${\rm Span}(\id,\tilde{u})\subset\tilde{\Oc}$. The commutant of this complex structure in ${\rm Spin}(11,3)$ is then ${\rm Spin}(7,1)\times {\rm Spin}(4,2)$, which does not seem to be interesting for physics. 

Let us now consider the case when $\tilde{u}$ is selected to be timelike. In this case the product of 6 $\gamma$-matrices that correspond to directions not in ${\rm Span}(\id,\tilde{u})\subset\tilde{\Oc}$ is not a complex structure, as this operator squares to plus the identity. However, now the product of the 4 $\gamma$-matrices that correspond to $\tilde{u}$ and to the 3 spacelike directions in ${\rm Cliff}_{4,3}$ is a complex structure. We shall refer to it as $\tilde{J}$. Its commutant in ${\rm Spin}(11,3)$ is ${\rm Spin}(10)\times{\rm Spin}(1,3)$, which is interesting. 

This discussion can be summarised by the following proposition:
\begin{theorem} There exist two complex structures $J, \tilde{J}$ on the space of semi-spinors of ${\rm Spin}(11,3)$, one parametrised by a unit imaginary octonion $u\in{\rm Im}(\Oc)$, the other parametrised by a unit imaginary octonion $\tilde{u}\in{\rm Im}(\tilde{\Oc})$ that is {\bf timelike}. Their common commutant in ${\rm Spin}(11,3)$ is the product of the Pati-Salam ${\rm Spin}(6)\times{\rm Spin}(4)$ and Lorentz ${\rm Spin}(1,3)$ groups. 
\end{theorem}
While it is perhaps not very surprising that complex structures breaking ${\rm Spin}(11,3)$ to Pati-Salam and Lorentz groups exist, we find it striking that both of these are parametrised by the same data -- a unit imaginary octonion in either $\Oc$ or $\tilde{\Oc}$. Further, the fact that there are two such natural complex structures is explained by the fact that there exists an $\Oc\otimes\tilde{\Oc}$ model for ${\rm Spin}(11,3)$. While this does not yet give a sought mechanism of symmetry breaking from ${\rm Spin}(11,3)$ to the groups that we see in Nature, the similarity of the complex structures $J, \tilde{J}$ suggests that maybe the dynamical mechanism that selects them is one and the same. In any case, the result in Theorem 2 puts the symmetry breaking needed to single out the Lorentz group from ${\rm Spin}(11,3)$ on a very similar footing with the symmetry breaking that is needed to select the Pati-Salam group from ${\rm Spin}(10)$ GUT group. This suggests that unconventional descriptions that "unify" the internal (colour, isospin) with the Lorentz spin degrees of freedom should be taken more seriously. 

The organisation of the rest of the paper is as follows. We start by reviewing in Section \ref{sec:spin7} how the group ${\rm Spin}(7)$ is described using the usual octonions. The spinor representation of ${\rm Spin}(7)$ is naturally the octonions $\Oc$. We also explain here how a natural complex structure $J$ on $\Oc$ arise from a choice of a unit imaginary octonion, and how the subgroups ${\rm SU}(3)$ and ${\rm U}(1)$ of ${\rm Spin}(7)$ arise in the process. We repeat similar analysis for ${\rm Spin}(4,3)$ in Section \ref{sec:spin43}. The novelty here is that a new type of complex structure arises, one related to a timelike unit octonion. The commutant of this complex structure $\tilde{J}$ is then the product of the Lorentz ${\rm Spin}(1,3)$ and "weak" ${\rm SU}(2)$ gauge groups. We put everything together in Section \ref{sec:spin11-3}, where the octonionic model for ${\rm Spin}(11,3)$ is described. We then interrupt our representation theoretic discussion by a review of some aspects of the left/right symmetric extension of the SM in Section \ref{sec:left-right}. This becomes a very useful starting point when we describe an explicit dictionary between particles and octonions in Section \ref{sec:dictionary}. 
We then apply the developed octonionic formalism to the problem of characterising the symmetry breaking and possible fermion mass terms. We make a small step in this direction and show how the  3-form field in $\Lambda^3(\R^{11,3})$ can be used to break the symmetry down to the product of the SM gauge group, Lorentz group, and ${\rm U}(1)_{B-L}$, which remains unbroken. This Higgs also produces the Dirac mass terms for all the particles. We conclude with a discussion.

\section{${\rm Spin}(7)$ and octonions}
\label{sec:spin7}

\subsection{Octonions}

Octonions, see e.g.,  \cite{Baez:2001dm}, are objects that can be represented as linear combinations of the unit octonions $1,e^1,\ldots, e^7$
\be
x = x_0 + x_1 e^1 +\ldots + x_7 e^7, \qquad x_0, x_1, \ldots, x_7 \in \R.
\ee
The conjugation is again the operation that flips the signs of all the imaginary coefficients
\be
\overline{x} = x_0 - x_1 e^1 -\ldots - x_7 e^7,
\ee
and we have 
\be
|x|^2=x\overline{x}= (x_0)^2 + (x_1)^2 + \ldots + (x_7)^2 \equiv |x|^2.
\ee
For later purposes, we note that if we represent an octonion as an 8-component column with entries $x_0,x_1,\ldots, x_7$, denoted by the same symbol $x$, we can write the norm as $|x|^2=x^T x$. 

Octonions $\Oc$ form a normed division algebra that satisfies the composition property $|xy|^2=|x|^2|y|^2$. The cross-products of the imaginary octonions $e^1,\ldots, e^7$ is most conveniently encoded into a 3-form in $\R^7$ that arises as 
\be
C(x,y,z)=\langle xy, z\rangle, \qquad x,y,z\in {\rm Im}(\Oc),
\ee
where the inner product $\langle\cdot,\cdot\rangle$ in $\Oc$ comes by polarising the squared norm
\be
\langle x,y\rangle =  {\rm Re}(x\overline{y}), \qquad x,y\in \Oc.
\ee
One convenient form of $C$ is
\be\label{C}
C = e^{567} + e^5\wedge (e^{41}-e^{23}) + e^6\wedge (e^{42}-e^{31}) + e^7\wedge (e^{43}-e^{12}),
\ee
where the notation is $e^{ijk}=e^i\wedge e^j\wedge e^k$. It is then easy to read off the products of distinct imaginary octonions from (\ref{C}). For example, $e^5 e^6=e^7$, which is captured by the first term in (\ref{C}). The particular form (\ref{C}) of the cross-product on ${\rm Im}(\Oc)$ is convenient because it manifests the ${\rm Spin}(3)\times{\rm Spin}(4)$ subgroups of ${\rm Spin}(7)$, and also one of the two maximal subgroups ${\rm Spin}(4)$ of the group of automorphisms of the octonions $G_2\subset{\rm Spin}(7)$ that preserves (\ref{C}). 

Octonions are non-commutative and non-associative, but alternative. The last property is equivalent to saying that any two imaginary octonions (as well as the identity) generate a subalgebra that is associative, and is a copy of the quaternion algebra $\Hq$.

\subsection{Clifford algebra ${\rm Cliff}_7$}

Octonions give rise to a very convenient model for the Clifford algebra ${\rm Cliff}_7$. Indeed, we can identify
\be
{\rm Cliff}_7 = {\rm Im}(\Oc).
\ee
To realise this identification, we consider the endomorphisms of $\Oc$ given by the operators 
\be
E^i := L_{e^i}
\ee 
of left multiplication by an imaginary octonion. These operators anti-commute and square to minus the identity
\be\label{cliff-rels}
E^i E^j  + E^j E^i = - 2\delta^{ij}.
\ee
Thus, the generate the Clifford algebra ${\rm Cliff}_7$. 

The Clifford generators can be described very explicitly as $8\times 8$ anti-symmetric matrices acting on 8-dimensional columns that represent elements of $\Oc$. Using the octonion product encoded in (\ref{C}) we get
 \begin{eqnarray}\label{Es}
& E^1 &= -E_{12}+E_{38}-E_{47}+E_{56} \\ \nonumber
& E^2 &= -E_{13}-E_{28}+E_{46}+E_{57} \\ \nonumber
& E^3 &= -E_{14}+E_{27}-E_{36}+E_{58} \\ \nonumber
& E^4 &= -E_{15}-E_{26}-E_{37}-E_{48} \\ \nonumber
& E^5 &= -E_{16}+E_{25}+E_{34}-E_{78} \\ \nonumber
& E^6 &= -E_{17}-E_{24}+E_{35}+E_{68} \\ \nonumber
& E^7 &= -E_{18}+E_{23}+E_{45}-E_{67} .
 \end{eqnarray}
 These are all $8\times 8$ real (anti-symmetric) matrices, and the octonions in the column are ordered as $\id,e^1,\ldots, e^7$. They also have the property $E^i \id = e^i$, where both $\id, e^i$ are 8-component columns. 

\subsection{Lie algebra ${\mathfrak so}(7)$}

With $8\times 8$ matrices $E^i$ giving the $\gamma$-matrices in 7 dimensions, the Lie algebra ${\mathfrak so}(7)$ is generated by the commutators of $\gamma$-matrices, or simply by the products of pairs of distinct $E^i$
\be\label{Lie-so7}
X_{\mathfrak so(7)} = \sum_{i<j} \omega_{ij} E^i E^j.
\ee
The matrix $X_{\mathfrak so(7)}$ is also anti-symmetric. This, in particular, implies that under the Lie algebra action $x\to X x$ the norm squared $x^T x$ is preserved. 

Given the explicit form (\ref{Es}) of the Clifford generators, it is easy to produce an explicit $8\times 8$ matrix representing a general Lie algebra element. It is most convenient to do such explicit calculations by matrix manipulations in Mathematica.

\subsection{Complex structure}

Any of the Clifford generators $E^i$ squares to minus the identity, and so is a complex structure on $\Oc$. Thus, a choice of a unit $u\in{\rm Im}(\Oc)$ gives rise to the complex structure $J := L_u$. Our octonion multiplication encoded in (\ref{C}) is such that $e^4$ is treated as a preferred element, and so we choose $u=e^4$ and $J= E^4$. 

It is interesting to describe the subalgebra of ${\mathfrak so}(7)$ that commutes with $J$. It is clear that all products $E^i E^j$ that do not contain $E^4$ commute with it. Thus, the subalgebra of ${\mathfrak so}(7)$ that commutes with $J$ is ${\mathfrak so}(6)$ that corresponds to rotations in the $e^{1,2,3, 5,6,7}$ plane. 

Given that this ${\mathfrak so}(6)$ commutes with a complex structure, it admits a complex description. In this description it is the Lie algebra of the group that acts by unitary transformations in e.g. the $(1,0)$ eigenspace of $J$. Indeed, the complex structure allows to identify $\Oc=\C^4$, and transformations commuting with the complex structure preserve the $(1,0)$ and $(0,1)$ eigenspaces of $J$. In our conventions the $(1,0)$ eigenspace is the one that corresponds to the eigenvalue $-\im$. All in all, we get ${\mathfrak so}(6)={\mathfrak su}(4)$. 

\subsection{A different complex structure}

There is a different choice of the complex structure on $\Oc$, which is also parametrised by a unit imaginary octonion. This leads to a different but very interesting commutant in ${\mathfrak so}(7)$. This has been described in \cite{Krasnov:2019auj}.

Instead of using the operator $L_u$ of left multiplication by a unit imaginary octonion, one can use the complex structure $J':= R_u$ given by the operator of the right multiplication $R_u$. This gives a complex structure, but one different from $L_u$. The operator $R_u$ agrees with $L_u$ on the copy of $\C$ spanned by $\id,u$, but acts with the opposite sign on the rest of $\Oc$. It is easy to check that the commutant of $R_u$ in ${\mathfrak so}(7)$ is the subalgebra ${\mathfrak su}(3)\times {\mathfrak u}(1)\subset {\mathfrak su}(4)$. Thus, this complex structure breaks ${\mathfrak so}(7)$ even further, and produces the very relevant for physics algebra ${\mathfrak su}(3)$ of the strong force and the "hypercharge" ${\mathfrak u}(1)$.\footnote{When we describe the left/right symmetric model below, it will become clear that the right interpretation of this ${\mathfrak u}(1)$ is that corresponding to ${\rm U}(1)_{B-L}$ symmetry.} 

We can also describe the $R_u$ complex structure in a way that strongly mimics what happens in the ${\mathfrak so}(4,3)$ setting of the next section. Let us introduce the operator $\rho$ that acts as the identity on ${\rm Span}(\id,u)$ and changes the sign of the 6-dimensional orthogonal to $\id,u$ space. In other words, let $\rho$ be the reflection in the $\id,u$ plane. This operator squares to $+\id$, and commutes with $L_u$. Their product is precisely $R_u$. Thus, we consider
\be\label{J-right}
J' := R_u = E^4 \rho.
\ee
As we already mentioned, the commutant of $J'$ in ${\mathfrak so}(7)$ is the direct sum ${\mathfrak su}(3)\oplus{\mathfrak u}(1)$.

\subsection{Parametrisation}

For later purposes, we now give a parametrisation of an octonion by its $(1,0)$ and $(0,1)$ coordinates. The $(1,0)$ coordinates of eigenvalue of $E^4$ of $-\im$ are given by
\be
L:= x_0-\im x_4, \quad Q_i = x_i -\im x_{i+4},
\ee
where we introduced suggestive names. An octonion $x$ with coordinates $x_0,x_1,\ldots, x_7$ is then parametrised as
\be
x_0 = \frac{1}{2}( L+ \bar{L}), \quad x_4= \frac{\im}{2}(L-\bar{L}), \quad x_i = \frac{1}{2}( Q_i +\bar{Q}_i), \quad x_{i+4} = \frac{\im}{2} ( Q_i - \bar{Q}_i).
\ee
Here the bar is both part of the name, as well as the complex conjugation that maps between the eigenspaces $(1,0), (0,1)$ of the complex structure $J$. Another complex structure and its related complex conjugation will appear later when we consider ${\rm Spin}(4,3)$. It will be important to distinguish them. 

For later purposes, we give formulas for various useful pairings in this parametrisation 
\be\label{O-formulas-1}
\langle x, x \rangle = \bar{L}L + \bar{Q}^i Q^i, \\ \nonumber
\langle x, (E^1-\im E^5)(E^2-\im E^6)(E^3-\im E^7) x \rangle= -4\im L^2, \\ \nonumber
\langle x, (E^1 E^5 + E^2 E^6+E^3 E^7)E^4 x \rangle = -3 \bar{L}L + \bar{Q}^i Q^i,
\ee
where the summation convention is implied. We will also need versions of these formulas with two different octonions paired. We have
\be\label{O-formulas-2}
\langle x_1, x_2 \rangle = \frac{1}{2}\left( \bar{L}_1 L_2 + \bar{L}_2 L_1  + \bar{Q}^i_1 Q^i_2 + \bar{Q}^i_2 Q^i_1\right), \\ \nonumber
\langle x_1, E^4 x_2 \rangle = \frac{\im}{2} \left( -\bar{L}_1 L_2 + \bar{L}_2 L_1  - \bar{Q}^i_1 Q^i_2 + \bar{Q}^i_2 Q^i_1\right), \\ \nonumber
\langle x_1, (E^1 E^5+E^2 E^6 +E^3 E^7) x_2 \rangle = \frac{\im}{2}\left( -3\bar{L}_1 L_2 + 3\bar{L}_2 L_1  - \bar{Q}^i_1 Q^i_2 + \bar{Q}^i_2 Q^i_1\right).
\ee
We will need these results when we compute the Yukawa mass terms. 

\subsection{The group of automorphisms of the octonions}

The group $G_2$ of automorphisms of the octonions is the subgroup of ${\rm SO}(7)$ that preserves the 3-form in (\ref{C}), and thus the cross-product. The 3-form (\ref{C}) can be written as
\be
C = - \sum_{i<j<k} \langle \id, E^i E^j E^j \id\rangle e^{ijk},
\ee
where the notation is that $e^{ijk}=e^i\wedge e^j\wedge e^k$. This shows that the stabiliser of $C$ in ${\mathfrak so}(7)$ is given by the transformations that fix the spinor corresponding to the identity octonion $\id$. An explicit calculation shows that this is the subalgebra of ${\mathfrak so}(7)$ satisfying
\be\label{g2}
-w_{27}+w_{36}-w_{45}=0, \quad w_{17}-w_{35}-w_{46}=0, \quad -w_{16}+w_{25}-w_{47}=0,
\\ \nonumber
-w_{14}-w_{23}+w_{67}=0, \quad w_{13}-w_{24}-w_{57}=0, \quad -w_{12}-w_{34}+w_{56}=0,
\ee
as well as $w_{15}+w_{26}+w_{37}=0$. This gives an explicit description of the Lie algebra ${\mathfrak g}(2)$. The dimension of this Lie algebra is 14.

\subsection{The group ${\rm SU}(3)$}

The special unitary group in 3 dimensions can be seen to arise in this context in many different ways. First, we can see it arising as the subgroup of ${\rm Spin}(7)$ that preserves two orthogonal spinors. If we take these to be $\id,u=e^4$, the subgroup that preserves $\id$ is $G_2$. Imposing the condition that $u$ is stabilised as well imposes 6 additional conditions $w_{i4}=0$, and produces the Lie algebra ${\mathfrak su}(3)$. 

A different, but equivalent way of seeing ${\rm SU}(3)$ arising is as the intersection of the group of automorphisms of the octonions $G_2$ with the group ${\rm Spin}(6)={\rm SU}(4)$ that commutes with $J$. Indeed, the conditions reducing to ${\mathfrak so}(6)={\mathfrak su}(4)$ are the 6 conditions $w_{i4}=0$. Intersected with (\ref{g2}) this gives us ${\mathfrak su}(3)$. 

Yet another way to see ${\rm SU}(3)$ is to note that while $J$ allows us to identify $\Oc=\C^4$, the unit imaginary octonion $u$ that was selected to produce this complex structure also gives us a preferred copy of the complex plane in $\C^4$, the one spanned by $\id,u$. The transformations from ${\rm SU}(4)$ do not preserve this copy of the complex plane, mixing all 4 directions in $\C^4$. The subgroup of ${\rm SU}(4)$ that preserves $\C={\rm Span}(\id,u)$ is ${\rm U}(3)$. The subgroup of this that fits into the group of automorphisms of the octonions is ${\rm SU}(3)$. 

The final remark we make here is that the ${\rm U}(1)_{B-L}\subset{\rm SU}(4)\subset{\rm Spin}(7)$ does not survive the intersection with the group of automorphisms of the octonions $G_2$. Indeed, its generator is precisely $w_{15}+w_{26}+w_{37}$, which is set to zero by the ${\mathfrak g}(2)$ conditions. This suggests that the dynamical mechanism that is to break the Pati-Salam ${\rm SU}(4)$ to ${\rm SU}(3)\times{\rm U}(1)_{B-L}$ is not one corresponding to taking the intersection with $G_2$. This remark will be important in the next section when we discuss the split analogs of all these statements. 

\section{${\rm Spin}(4,3)$ and the split octonions}
\label{sec:spin43}

\subsection{Split octonions}

Split octonions $\tilde{\Oc}$ are similarly generated by the identity and seven imaginary octonions
\be
y= y_0 + y_1 \tilde{e}^1 +\ldots + y_7 \tilde{e}^7.
\ee
To distinguish the split from the usual octonions we will denote them either with a letter with a tilde, or use letter $y$ (with an index if necessary) rather than $x$. The unit split octonions satisfy
\be
(\tilde{e}^{5,6,7})^2=-1, \qquad (\tilde{e}^{1,2,3,4})^2=1.
\ee
The split octonion conjugation again changes the sign of all the imaginary units. This means that the split octonion quadratic form
\be\label{Q-split}
|y|^2_s:= y\bar{y} = (y_0)^2 - (y_1)^2 - (y_2)^2 - (y_3)^2 - (y_4)^2 + (y_5)^2 + (y_6)^2 + (y_7)^2
\ee
is of the split signature $(4,4)$. It will be convenient to rewrite this quadratic form in a matrix notation. We introduce a diagonal matrix $\tau = {\rm diag}(-1,1,1,1,1,-1,-1,-1)$, which described a reflection in the $1,2,3,4$ plane. If we represent a split octonion by a 8-dimensional column with entries $y_0,y_1,\ldots, y_7$, the quadratic form is
\be\label{norm-split}
|y|^2 = -y^T \tau y.
\ee
This form of writing will be useful later.

The split octonion cross-product is again most conveniently encoded by a 3-form, which we choose to be
\be\label{C-split}
C = \tilde{e}^{567} - \tilde{e}^5\wedge (\tilde{e}^{41}-\tilde{e}^{23}) - \tilde{e}^6\wedge (\tilde{e}^{42}-\tilde{e}^{31}) - \tilde{e}^7\wedge (\tilde{e}^{43}-\tilde{e}^{12}).
\ee
Note that the only change as compared to (\ref{C}) is that the signs in front of the last 3 terms changed. This means that the cross-product in the $\tilde{e}^{5,6,7}$ plane is unchanged to the case of usual octonions. This plane, together with $\id$, generates a copy of the quaternions $\Hq$. The signs of all other cross-products get reversed as compared to $\Oc$. 

\subsection{The Clifford algebra ${\rm Cliff}_{4,3}$}

Operators $\tilde{E}^i := L_{ \tilde{e}^i}$ of left multiplication by a unit split imaginary octonion generate the Clifford algebra ${\rm Cliff}_{4,3}$
\be
{\rm Cliff}_{4,3} = {\rm Im}(\tilde{\Oc}).
\ee
We can represent them explicitly as the following $8\times 8$ matrices
\begin{eqnarray}\label{E-split}
& \tilde{E}^1 &= S_{12}+S_{38}-S_{47}+S_{56} \\ \nonumber
& \tilde{E}^2 &= S_{13}-S_{28}+S_{46}+S_{57} \\ \nonumber
& \tilde{E}^3 &= S_{14}+S_{27}-S_{36}+S_{58} \\ \nonumber
& \tilde{E}^4 &= S_{15}-S_{26}-S_{37}-S_{48} \\ \nonumber
& \tilde{E}^5 &= -E_{16}+E_{25}+E_{34}-E_{78} \\ \nonumber
& \tilde{E}^6 &= -E_{17}-E_{24}+E_{35}+E_{68} \\ \nonumber
& \tilde{E}^7 &= -E_{18}+E_{23}+E_{45}-E_{67} .
 \end{eqnarray}
 These satisfy the Clifford algebra relations
 \be
 \tilde{E}^i \tilde{E}^j + \tilde{E}^j \tilde{E}^i = 2\eta^{ij},
 \ee
 where $\eta={\rm diag}(1,1,1,1,-1,-1,-1)$ is the metric of signature $(4,3)$. Note that we have reversed the signs here as compared to (\ref{cliff-rels}), this will be convenient later when we put $\Oc$ and $\tilde{\Oc}$ together. This extra sign is the same as the one required in (\ref{signature}). The matrices $S_{ij}$ are symmetric matrices with the identity in $i$th row and $j$th column, as well as in $j$th row and $i$th column. It is then clear that the matrices $\tilde{E}^{1,2,3,4}$ are symmetric, while $\tilde{E}^{5,6,7}$ are anti-symmetric. Note that $\tilde{E}^{5,6,7}=E^{5,6,7}$. Again we have the property that $\tilde{E}^i \tilde{\id} = \tilde{e}^i$, where both $\tilde{\id}, \tilde{e}^i$ are viewed as 8-component columns. 
 
 \subsection{Lie algebra ${\mathfrak so}(4,3)$}
 
 Forming the products of pairs of distinct Clifford generators we get the representation of a general Lie algebra in terms of $8\times 8$ matrices
 \be\label{Lie-so43}
 X_{{\mathfrak so}(4,3)} = \sum_{i<j} \tilde{\omega}_{ij} \tilde{E}^i \tilde{E}^j.
 \ee
These matrices no longer have a definite symmetry. Instead, the invariance of the quadratic form (\ref{norm-split}) can be written as the property 
\be
X_{{\mathfrak so}(4,3)}^T \tau + \tau X_{{\mathfrak so}(4,3)}=0.
\ee
Here $\tau$ is the matrix of the quadratic form (\ref{norm-split}).

\subsection{Complex structures}

We now enter into a less familiar part of the discussion. It is clear that any of the 3 Clifford generators $\tilde{E}^{5,6,7}$ (or any unit vector constructed as their linear combination) can serve as a complex structure on $\tilde{\Oc}$. This complex structure identifies $\tilde{\Oc}=\C^4$. What commutes with it in ${\mathfrak so}(4,3)$ is the subalgebra ${\mathfrak so}(4,2)$. But this is not the complex structure that is of interest to us in relation to physics.

Any of the generators $\tilde{E}^{1,2,3,4}$ squares to plus the identity, and so is not a complex structure. But we also have the object
\be
\tau = \tilde{E}^5 \tilde{E}^6 \tilde{E}^7.
\ee
This is an operator that reverses the signs of the positive-definite directions $\tilde{\id}, \tilde{e}^{5,6,7}$ of the split quadratic form (\ref{Q-split}), and leaves the negative-definite directions $\tilde{e}^{1,2,3,4}$ intact. As a matrix, this is the already encountered matrix of the quadratic form (\ref{norm-split}). This operator squares to plus the identity $\tau^2=\id$, but {\bf anti-commutes} with any of the Clifford generators $\tilde{E}^{1,2,3,4}$ (and commutes with all the $\tilde{E}^{5,6,7}$ generators). Thus, the product of $\tau$ with any one of $\tilde{E}^{1,2,3,4}$ (or with a unit linear combination constructed from them) is a complex structure. We choose $\tilde{u}=\tilde{e}^4$ and define
\be
\tilde{J} := \tilde{E}^4 \tau.
\ee
Note the strong analogy with (\ref{J-right}). This is the complex structure that is of our main interest. Indeed, being given by the product of 4 Clifford generators, its commutant in ${\mathfrak so}(4,3)$ is the algebra ${\mathfrak so}(3)\times{\mathfrak so}(1,3)$. Both groups are very interesting, because the first factor reminds us of the ${\rm SU}(2)$ acting on isospin of particles, while the second factor is the Lorentz group. The first factor is the one describing rotations in the $\tilde{e}^{1,2,3}$ plane, while the second one mixes the directions $\tilde{e}^{4,5,6,7}$. The direction $\tilde{e}^4$ is a timelike direction with respect to the split octonion quadratic form (\ref{Q-split}), and we will refer to it as such in what follows. For later purposes we note that we can rewrite the complex structure $\tilde{J}$ as
\be\label{J-split-123}
\tilde{J} := -\tilde{E}^1 \tilde{E}^2\tilde{E}^3,
\ee
i.e., as the product of the timelike Clifford generators that are distinct from the chosen $\tilde{E}^4$. 

\subsection{Decomposition of $\tilde{\Oc}$ under ${\mathfrak so}(3)\times{\mathfrak so}(1,3)$}

The split octonions $\tilde{\Oc}$ form the spinor representation of ${\rm Spin}(4,3)$. When we restrict to ${\rm Spin}(3)\times{\rm Spin}(1,3)$, we expect the spinor to transform as spinor with respect to both of the factors. It is nevertheless very interesting to see how this happens explicitly. 

The $(1,0)$ coordinates on $\tilde{\Oc}$ for the complex structure $\tilde{J}$ are given by
\be
w_0:= \frac{1}{2}(y_0 + \im y_4), \quad w_i:= \frac{1}{2}(y_i +\im y_{i+4}), 
\ee
where the index $i=1,2,3$ and we introduced the factors of $1/2$ for future convenience. 
The transformations from ${\mathfrak so}(3)\times{\mathfrak so}(1,3)$ commute with $\tilde{J}$ and thus preserve the $(1,0)$ subspace. Therefore, they can be described as $4\times 4$ complex matrices acting on the 4-columns $(w_0, w_1, w_2, w_3)$. Explicitly, we have
\be
X_{{\mathfrak so}(3)} = \left( \begin{array}{cccc} 0 & \im x & \im y & \im z \\
\im x & 0 &  z&  -y  \\
\im y &  - z & 0 & x \\
\im z & y & - x & 0 \end{array}\right), 
\ee
where
\be 
x= \tilde{\omega}_{23}, \quad y =  -\tilde{\omega}_{13}, \quad z= \tilde{\omega}_{12}
\ee
are 3 real parameters. We also have 
\be
X_{{\mathfrak so}(1,3)} = \left( \begin{array}{cccc} 0 & a & b& c \\
a& 0 &  \im c &  -\im b  \\
b &  - \im c & 0 & \im a \\
c& \im b & - \im a & 0 \end{array}\right),
\ee
where
\be
a= -\tilde{\omega}_{45} + \im \tilde{\omega}_{67}, \quad b= -\tilde{\omega}_{46} - \im \tilde{\omega}_{57}, \quad c=-\tilde{\omega}_{47} + \im \tilde{\omega}_{56}
\ee
are complex parameters. It can be verified that the matrices $X_{{\mathfrak so}(3)}$ and $X_{{\mathfrak so}(1,3)}$ commute. 

It is then easy to check that the 2-component columns
\be\label{ud}
u = \left(\begin{array}{c} u_1 \\ u_2 \end{array}\right):= \left( \begin{array}{c} w_0 + w_3 \\ w_1-\im w_2 \end{array}\right), \qquad
d= \left(\begin{array}{c} d_1 \\ d_2 \end{array}\right):= \left( \begin{array}{c} w_1+\im w_2 \\ w_0 - w_3 \end{array}\right)
\ee
transform under $X_{{\mathfrak so}(1,3)}$ transformations as 2-component spinors of the same Lorentz chirality, that is
\be
u\to A u, \qquad d\to Ad, \qquad 
A=\left( \begin{array}{cc} c & a+\im b \\ a-\im b & -c \end{array}\right).
\ee
On the other hand, the 2-component column with $u, d$ as entries transforms under $X_{{\mathfrak so}(3)}$ as
\be
\left(\begin{array}{c} u \\ d \end{array}\right) \to \left(\begin{array}{cc} \im z & y +\im x \\ - y+\im x & -\im z\end{array}\right) \left(\begin{array}{c} u \\ d \end{array}\right).
\ee
Summarising, by introducing the complex structure $\tilde{J}$, the space of split octonions $\tilde{\Oc}=\R^8$ splits into its $(1,0)$ and $(0,1)$ eigenspaces. The $(1,0)$ eigenspace $\C^4$ transforms as the spinor representation of ${\rm Spin}(3)$, as well as the 2-component spinor representation of the Lorentz group ${\rm Spin}(1,3)$
\be
\C^4_{(1,0)} = ( {\bf 2}, {\bf 2}, {\bf 1}).
\ee
The complex conjugate $(0,1)$ eingespace transforms as the spinor of ${\rm Spin}(3)$, as well as the complex conjugate 2-component spinor representation of Lorentz
\be
\C^4_{(0,1)} = ( {\bf 2}, {\bf 1}, {\bf 2}).
\ee
Thus, a single copy of $\tilde{\Oc}$ gives us a 2-component Lorentz spinor that is at the same time a spinor of what can be identified as "weak" ${\rm SU}(2)$.\footnote{Below we will see that this ${\rm SU}(2)$ corresponds to the diagonal subgroup in ${\rm SU}(2)_L\times{\rm SU}(2)_R$.} Thus, the "weak" ${\rm SU}(2)$ and the Lorentz group are very naturally unified within ${\rm Spin}(4,3)$. 

\subsection{Parametrisation}

It will help if we develop the notation a bit further. We have seen that a ${\rm Spin}(4,3)$ spinor $y\in\tilde{\Oc}$ splits into its $(1,0), (0,1)$ components
\be
y = \tilde{P}_+ y + \tilde{P}_- y := y^+ + y^-,
\ee
where
\be
\tilde{P}_\pm = \frac{1}{2} ( \id \pm \im \tilde{J})
\ee
are the projectors. The split octonion $y^+$ transforms as a 2-component Lorentz and ${\rm SU}(2)$ spinors, and $y^-$ transforms as a 2-component Lorentz spinor of the opposite chirality, and again as a ${\rm SU}(2)$ spinor. Note that the two complex octonions $y^\pm$ are related by the complex conjugation $y^-= (y^+)^*$.

It will be helpful to write down the explicit parametrisation of the octonions $y_\pm$ by the components of the Lorentz and ${\rm SU}(2)$ spinors. Using (\ref{ud}) we have
\be
w_0 = \frac{1}{2}( u_1+d_2), \quad w_3=\frac{1}{2}( u_1-d_2), \quad
w_1=\frac{1}{2}( u_2+d_1), \quad w_2=\frac{\im}{2}( u_2-d_1).
\ee
We also have for the real coordinates on $\tilde{\Oc}$
\be
y_0 =  (w_0 + w_0^*), \quad y_4=-\im(w_0-w_0^*), \quad
y_i = ( w_i + w_i^*), \quad y_{i+4} = -\im( w_i - w_i^*),
\ee
with $i=1,2,3$, and where the star denotes the complex conjugation. This means that we have
\be
y^+ = \frac{1}{2} \left( \begin{array}{c} u_1+d_2 \\ u_2 + d_1 \\ \im (u_2-d_1) \\ u_1-d_2 \\ 
-\im (u_1+d_2) \\ -\im (u_2+d_1) \\ (u_2-d_1) \\ -\im (u_1-d_2) \end{array}\right).
\ee

If we introduce an octonion $y$ parametrised in this way, a computation gives the expected Lorentz and ${\rm SU}(2)$ invariant pairing
\be
\langle y, y \rangle =  \langle y^+, y^+ \rangle +  \langle y^-, y^- \rangle= \left( \begin{array}{cc} u & d \end{array}\right) 
\left( \begin{array}{cc} 0 & 1 \\ -1 & 0 \end{array}\right)  \left( \begin{array}{c} u \\ d \end{array}\right) + {\rm c.c.},
\ee
where c.c. stands for "complex conjugate", and we have used the index-free 2-component spinor notation for the pairing of 2-component spinors
\be
ud = -du :=u^T \epsilon d, \qquad u= \left( \begin{array}{c} u_1 \\ u_2 \end{array}\right) , d=\left( \begin{array}{c} d_1 \\ d_2 \end{array}\right), \quad  \epsilon:= \left( \begin{array}{cc} 0 & 1 \\ -1 & 0 \end{array}\right).
\ee
From now on, we will never spell out the Lorentz pairing of 2-component spinors in terms of matrices, always using the index-free (and matrix-free) notation. But in order to avoid confusion, it will be useful to explicitly write the ${\rm SU}(2)$ invariant pairing in matrix terms. Also, from now on "complex conjugate" will always refer to the operation that maps between the $(1,0), (0,1)$ eigenspaces of the complex structure $\tilde{J}$. When we consider ${\rm Spin}(11,3)$, there will be another complex structure floating around, but in order to avoid confusion we will always spell out the complex conjugate with respect to this other complex structure explicitly. 

We can also write a more general formula for the pairing of two different ${\rm Spin}(4,3)$ spinors. We have
\be\label{y12}
\langle y_1, y_2\rangle = \langle y_1^+, y_2^+ \rangle + \langle y_1^-, y_2^- \rangle =
\left( \begin{array}{cc} u & d \end{array}\right)_1 
\left( \begin{array}{cc} 0 & 1 \\ -1 & 0 \end{array}\right)  \left( \begin{array}{c} u \\ d \end{array}\right)_2 + {\rm c.c.}
\ee
The first term is a Lorentz and ${\rm SU}(2)$ invariant pairing of two different 2-component Lorentz and ${\rm SU}(2)$ spinors. 

\subsection{Some useful formulas}

Here we list some useful pairings that can be computed using the particle-friendly parametrisation developed in the previous subsection. We will give them in their version analogous to (\ref{y12}), with two different spinors. 

First, we have
\be\label{Ot-formulas-1}
- \langle y_1, \tilde{E}^1\tilde{E}^2 \tilde{E}^3 y_2 \rangle = \langle y_1, \tilde{J} y_2 \rangle=
-\im \left( \begin{array}{cc} u & d \end{array}\right)_1
\left( \begin{array}{cc} 0 & 1 \\ -1 & 0 \end{array}\right)  \left( \begin{array}{c} u \\ d \end{array}\right)_2 + {\rm c.c.}
\ee
This result follows from the fact that $\tilde{J}$ acts as the operator of multiplication by $-\im$ on $(1,0)$ eigenspace where $y_{1,2}^+$ take values. 

We will also need
\be\label{Ot-formulas-2}
\langle y_1, (\phi^i \tilde{E}^i) y_2\rangle = \left( \begin{array}{cc} u & d \end{array}\right)_1 
\left( \begin{array}{cc} 0 & 1 \\ -1 & 0 \end{array}\right) \left( \begin{array}{cc} \phi^3 & \phi^1 -\im \phi^2 \\ \phi^1+\im\phi^2 & -\phi^3 \end{array}\right) \left( \begin{array}{c} u \\ d \end{array}\right)_2 + {\rm c.c.}
\\ \nonumber
\langle y_1, \frac{1}{2}(\epsilon^{ijk} \phi^i \tilde{E}^j \tilde{E}^k) y_2\rangle = \im \left( \begin{array}{cc} u & d \end{array}\right)_1 
\left( \begin{array}{cc} 0 & 1 \\ -1 & 0 \end{array}\right) \left( \begin{array}{cc} \phi^3 & \phi^1 -\im \phi^2 \\ \phi^1+\im\phi^2 & -\phi^3 \end{array}\right) \left( \begin{array}{c} u \\ d \end{array}\right)_2 + {\rm c.c.}
\ee

\subsection{The split $\tilde{G}_2$}

The group of automorphisms of the split octonions $\tilde{G}_2$ can be defined as the subgroup of ${\rm SO}(4,3)$ that preserves the 3-form (\ref{C-split}). Equivalently, it is the subgroup of ${\rm Spin}(4,3)$ that preserves the spinor $\id$. 

It can be checked that the intersection of the Lie algebras $\tilde{{\mathfrak g}}(2)$ and ${\mathfrak so}(3)\times{\mathfrak so}(1,3)$ is the diagonal ${\mathfrak so}(3)$ that describes simultaneous rotations in the $\tilde{e}^{1,2,3}$ and $\tilde{e}^{5,6,7}$ planes. In particular, there are no boosts in this intersection. 

While this may sound as bad news for physics, we remind the reader that the analogous intersection in the $\Oc$ case is that between ${\mathfrak g}(2)$ and ${\mathfrak so}(6)$. And as we discussed in the previous section, this intersection is ${\mathfrak su}(3)$, but not ${\mathfrak su}(3)\times {\mathfrak u}(1)_{B-L}$ that would be desirable for physics. So, the ${\rm U}(1)_{B-L}$ does not survive the intersection with $G_2$ in the case of ${\rm Spin}(7)$. What we see in the present case of ${\rm Spin}(4,3)$ is that this group naturally contains the weak ${\rm SU}(2)$ together with the Lorentz group, but neither of this survives the intersection with $\tilde{G}_2$. Only the diagonal ${\rm SU}(2)$ consisting of the simultaneous weak ${\rm SU}(2)$ transformations and spatial rotations does survive as a subgroup of $\tilde{G}_2$. So, this situation is analogous to what happens for ${\rm Spin}(7)$. This just means that the to-be-found mechanism that will select the "physical" unbroken subgroups from ${\rm Spin}(11,3)$ is not the one where intersections with $G_2, \tilde{G}_2$ will play role. But there is clearly something important to be understood here. 

\section{The octonionic model for ${\rm Spin}(11,3)$}
\label{sec:spin11-3}

We now put the $\Oc, \tilde{\Oc}$ building blocks together, and construct an octonionic model for the group ${\rm Spin}(11,3)$. The construction we are to describe applies to a more general set of pseudo-orthogonal groups, and so we give it in full generality. 

\subsection{Yet another magic square}

There is an octonionic model for the exceptional Lie groups, which is obtained by considering $3\times 3$ matrices with entries in $\X\otimes\Y$, where $\X,\Y$ are division algebras, see \cite{Barton:2000ki}. This leads to what in the literature is known as the "magic square" construction. These authors also describe the $2\times 2$ matrix version of the magic square. This gives a set of octonionic models for groups the largest of which is ${\rm Spin}(12,4)$. We describe a very closely related construction, which effectively forgets two of the corresponding Clifford generators, to produce models with the largest covered group being ${\rm Spin}(11,3)$. The difference between our construction and that in \cite{Barton:2000ki}, see also \cite{Dray:2014mba}, is that in our case the semi-spinor is the space $\X\otimes\Y$, while in the former case the semi-spinor is twice that $(\X\otimes\Y)^2$.

We now work out a table of cases that are covered by the construction of Theorem 1. There are two tables that can be produced. One corresponds to taking $\X$ to be a division algebra, and $\Y$ to be a split (i.e. just composition) algebra. This produces the most interesting "magic square". The largest group covered by this construction is ${\rm Spin}(11,3)$.  
\be
\label{table}
\begin{tabular}{c | ccc}
 & $\C$ & $\Hq$ & $\Oc$ \\
 \hline
 $\tilde{\C}$ & (2,0) & (4,0) & (8,0) \\
 $\tilde{\Hq}$ & (3,1) & (5,1) & (9,1) \\
 $\tilde{\Oc}$ & (5,3) & (7,3) & (11,3)
 \end{tabular}\qquad
 \begin{tabular}{c | ccc}
 & $\C$ & $\Hq$ & $\Oc$ \\
 \hline
 $\C$ & (1,1) & (3,1) & (7,1) \\
 $\Hq$ & (1,3) & (3,3) & (7,3) \\
 $\Oc$ & (1,7) & (3,7) & (7,7)
 \end{tabular}
 \ee
 What is indicated in these tables is the signature of the arising pseudo-orthogonal group. 
 
 The second table is the case when $\X,\Y$ are either both division, or both split. These two possibilities  give the same groups. The diagonal in this case gives the split signature pseudo-orthogonal groups ${\rm Spin}(1,1), {\rm Spin}(3,3)$ and ${\rm Spin}(7,7)$. The only other case that is not covered by the first table is that of ${\rm Spin}(7,1)$. 
 
 It is clear that adding $(1,1)$ to every entry of the first table reproduces the $2\times 2$ tables in  \cite{Barton:2000ki}, \cite{Dray:2014mba}. However, it is not so easy to go from the construction of Theorem 1 to that in \cite{Barton:2000ki}, \cite{Dray:2014mba} because these references describe a model for the Lie algebra of the relevant groups as the algebra of $2\times 2$ matrices with entries in $\X\otimes \Y$. The corresponding $\gamma$-matrices are then $4\times 4$ such matrices. So, giving a model for the Clifford algebra requires one higher level of complexity. This can be done, but is not relevant for our purposes. We refer to \cite{Dray:2014mba} for details. 

\subsection{Lie algebra}

Taking the commutators of the $\Gamma$-matrices (\ref{Gammas}), or simply the products of distinct $\Gamma$-matrices, we get the following description of the Lie algebra of the groups that our construction covers. The Lie algebra is given by the following endomorphisms of $\X\otimes \Y$
\be
{\mathfrak so}({\rm Im}(\X))\otimes \id + \id\otimes {\mathfrak so}({\rm Im}(\Y)) + L_x \otimes L_y, \qquad x\in{\rm Im}(\X), y\in{\rm Im}(\Y).
\ee
For the case of ${\mathfrak so}(11,3)$ that is of most interest for us we can give a more explicit description
\be\label{lie-11-3}
X_{{\mathfrak so}(11,3)}= \sum_{i<j} \omega_{ij} E^i E^j \otimes \id + \id\otimes \sum_{i<j} \tilde{\omega}_{ij} \tilde{E}^i \tilde{E}^j  + \sum_{i,j} a_{ij} E^i \otimes \tilde{E}^j.
\ee
Thus, there is the Lie algebra ${\mathfrak so}(7)$ acting on the $\Oc$ factor in $\Oc\otimes\tilde{\Oc}$, the algebra ${\mathfrak so}(4,3)$ acting on the $\tilde{\Oc}$ factor, and the terms that mix the first and second sets of directions in $(11,3)=(7,0) + (4,3)$. All the blocks here are constructed from $8\times 8$ matrices $E^i,\tilde{E}^i$, see (\ref{Es}), (\ref{E-split}). This gives a very explicit description of the Lie algebra. 

\subsection{Invariant pairing}

There is no ${\mathfrak so}(11,3)$ invariant pairing on the space of semi-spinors $S_+$. This is because the matrices appearing in (\ref{lie-11-3}) have different symmetry properties. The first two terms are anti-symmetric (in appropriate sense)
\be
(\omega_{ij} E^i E^j)^T = \omega_{ij} E^j E^i = - \omega_{ij} E^i E^j, 
\ee
and, using $(\tilde{E}^i)^T \tau = - \tau \tilde{E}^i$
\be
(\tilde{\omega}_{ij} \tilde{E}^i \tilde{E}^j)^T \tau = \tau \tilde{\omega}_{ij} \tilde{E}^j \tilde{E}^i = - \tau \tilde{\omega}_{ij} \tilde{E}^i \tilde{E}^j .
\ee
The last term is, however, symmetric
\be
(a_{ij} E^i \otimes \tilde{E}^j)^T (\id\otimes \tau) = (\id\otimes \tau) a_{ij} E^i \otimes \tilde{E}^j.
\ee
Thus, if we build a paring on $\Oc\otimes\tilde{\Oc}$ using the octonion pairings we have on each of these spaces, it will be invariant under ${\mathfrak so}(7),{\mathfrak so}(4,3)$, but not under ${\mathfrak so}(11,3)$.

Instead, there exists an invariant pairing between the spaces of different types of semi-spinors $S_\pm$. The space $S_-$ can also be identified with $\Oc\otimes \tilde{\Oc}$. For the expression (\ref{Gammas}) for the $\Gamma$-matrices it is easy to see that the action of the Lie algebra on $S_-$ is given by
\be
X'_{{\mathfrak so}(11,3)}= \sum_{i<j} \omega_{ij} E^i E^j \otimes \id + \id\otimes \sum_{i<j} \tilde{\omega}_{ij} \tilde{E}^i \tilde{E}^j  - \sum_{i,j} a_{ij} E^i \otimes \tilde{E}^j.
\ee
Note the different sign in front of the last term as compared to (\ref{lie-11-3}). If we denote elements of $S_+=\Oc\otimes\tilde{\Oc}$ by $\Psi$, and those of $S_-=\Oc\otimes \tilde{\Oc}$ by $\Psi'$ we have the following invariant pairing
\be
\lla \Psi', \Psi\rra,
\ee
where the double angle brackets denote the composition of the pairings on $\Oc$ and $\tilde{\Oc}$. The ${\mathfrak so}(11,3)$ invariance is the easily checked property
\be
\lla  X'_{{\mathfrak so}(11,3)} \Psi', \Psi\rra + \lla \Psi', X_{{\mathfrak so}(11,3)} \Psi\rra=0.
\ee

\section{The left/right symmetric models}
\label{sec:left-right}

We now interrupt our representation theoretic discussion to review the left/right symmetric extensions of the SM, as these are most closely related to our formalism. We follow \cite{DiLuzio:2011mda}, see in particular Section 1.3.1.

\subsection{Spinor fields}

It is convenient to organise the fermionic fields into the following multiplets
\be\label{fermions}
Q^i = \left( \begin{array}{c} u^i \\ d^i \end{array}\right), \quad
L = \left( \begin{array}{c} \nu \\ e \end{array}\right), \quad
\bar{Q}^i = \left( \begin{array}{ccc} \bar{d}^i \\ -\bar{u}^i \end{array}\right), \quad
\bar{L} = \left( \begin{array}{c} \bar{e} \\ -\bar{\nu} \end{array}\right).
\ee
These fields transform under the left/right-symmetric gauge group 
\be\label{GLR}
G_{\rm LR}={\rm SU}(3)\times{\rm SU}(2)_L\times{\rm SU}(2)_R\times {\rm U}(1)_{B-L}
\ee
 in the following representations
\be\label{charges}
Q = (3,2,1,+\frac{1}{3}), \quad L= (1,2,1,-1), \quad \bar{Q}=(\bar{3}, 1, 2, -\frac{1}{3}), \quad \bar{L}= (1,1,2,+1).
\ee
Specifically, the transformation properties under ${\rm SU}(2)_L\times{\rm SU}(2)_R$ are
\be
Q\to U_L Q, \quad L\to U_L L, \quad \bar{Q} \to U_R \bar{Q}, \quad \bar{L}\to U_R \bar{L}.
\ee

\subsection{Bi-doublet Higgs field}

Most (but not all, see below) versions of the left/right symmetric model require a Higgs field in the bi-doublet representation 
\be\label{Phi}
\Phi = (1,2,2,0), \qquad \Phi \to U_L \Phi U_R^\dagger.
\ee
Out of the components of $\Phi^*$ of the complex conjugate matrix one can construct another matrix
\be\label{epsilon}
\tilde{\Phi} = \epsilon \Phi^* \epsilon^T, \qquad \epsilon:=\left( \begin{array}{cc} 0 & 1 \\ -1 & 0\end{array}\right)
\ee
that has the same transformation properties $\tilde{\Phi}\to U_L \tilde{\Phi} U_R^\dagger$ as $\Phi$.
The VEV of this Higgs field
\be\label{Phi-vev}
\Phi =\left( \begin{array}{cc} v_1 & 0 \\ 0 & v_2 \end{array}\right), \qquad
\tilde{\Phi} =\left( \begin{array}{cc} v^*_2 & 0 \\ 0 & v^*_1 \end{array}\right)
\ee
breaks the symmetry down to the diagonal ${\rm U}(1)\subset{\rm SU}(2)_L\times{\rm SU}(2)_R$. The ${\rm U}(1)_{B-L}$ remains unbroken. To break the remaining symmetries to the electromagnetic ${\rm U}(1)$ one needs to introduce further Higgs fields, see below.

It is very convenient to have the Higgs field $\Phi$ because it allows construction of Yukawa mass terms. The $G_{\rm LR}$ invariant (lepton) mass term that can be constructed from $\Phi,\tilde{\Phi}$ is
\be\label{Phi-Y}
Y_L L^T \epsilon^T \Phi \bar{L} + \tilde{Y}_L L^T \epsilon^T \tilde{\Phi} \bar{L},
\ee
where $Y_L,\tilde{Y}_L$ are mass parameters (matrices in the case there is more than one generation). Evaluating these terms on the VEV's of $\Phi,\tilde{\Phi}$ we get
\be
(Y_L v_1 +\tilde{Y}_L v_2^*) e\bar{e} + (Y_L v_2 +\tilde{Y}_L v_1^*) \nu\bar{\nu},
\ee
which are the Dirac mass terms for the electron and the neutrino. Note that it is sufficient to take $v_1\not=0, v_2=0$ to obtain the Dirac masses. 

\subsection{Two Higgs in the adjoint}

As we already said, the Higgs field $\Phi$ cannot break the ${\rm U}(1)_{B-L}$ symmetry. One thus needs additional Higgs fields. The option that currently seems preferred in the literature goes under the name of the minimal left/right symmetric model, see e.g. \cite{Maiezza:2016ybz}. It introduces two more Higgs fields in the adjoint of each ${\rm SU}(2)_{L,R}$. It also requires neutrinos to be Majorana particles, and provides a natural room for the seesaw mechanism. The adjoint Higgs fields are $(B-L)$ charged and transform in the following representations
\be
\Delta_L = (1,3,1,+2), \quad \Delta_R=(1,1,3,+2),
\ee
with ${\rm SU}(2)_L\times{\rm SU}(2)_R$ transformation properties
\be
\Delta_L\to U_L \Delta_L U_L^\dagger, \qquad \Delta_R\to U_R \Delta_R U_R^\dagger.
\ee
The required VEV's for these fields are
\be
\Delta_{L,R} = \left(\begin{array}{cc} 0 & 0 \\ v_{L,R} & 0 \end{array}\right).
\ee
The field $\Delta_L$ leaves unbroken the diagonal subgroup in ${\rm U}(1)_L \times{\rm U}(1)_{B-L}$, analogously for $\Delta_R$. This means that there is a single electromagnetic ${\rm U}(1)_{EM}$ unbroken. The electric charge of all the fields is worked out via
\be
Q= T^3_L + T^3_R + \frac{B-L}{2}.
\ee
Thus, the hypercharge $Y=T^3_R + (B-L)/2$.

The adjoint Higgs fields allow for the construction of Yukawa mass terms. The lepton sector mass terms that can be written using $\Delta_{L,R}$ are
\be
Y_\Delta ( L^T \epsilon \Delta_L L + \bar{L}^T  \Delta^*_R \epsilon \bar{L}),
\ee
where $Y_\Delta$ is a mass parameter (a mass matrix in the case when there is more than one generation). This term is designed to be invariant under the discrete $\Z_2$ symmetry 
\be
L \leftrightarrow \bar{L}, \qquad \Delta_L \leftrightarrow \Delta_R^\dagger.
\ee
Evaluating everything on the VEV we get the following contribution to leptonic mass terms 
\be
 Y_\Delta(v_L \nu \nu + v_R^* \bar{\nu}\bar{\nu}).
\ee
These are the neutrino Majorana mass terms giving rise to the seesaw mechanism. 

\subsection{Two Higgs fields in the fundamental}

Another option is to add two Higgs fields in the fundamental representations of ${\rm SU}(2)_{L,R}$. This is the original left/right symmetric model studied in \cite{Senjanovic:1978ev}. But there is no natural realisation of the seesaw mechanism. 

Let us introduce the two Higgs fields $\chi_{L,R}$ transforming as under $G_{\rm LR}$ as
\be
\chi_L = (2,1,+1), \qquad \chi_R=(1,2,+1).
\ee
There exists a vacuum configuration in which $\Phi$ takes the form (\ref{Phi-vev}) and 
\be\label{fund-H-vev}
\chi_L=\left( \begin{array}{c} v_L \\  0 \end{array}\right), \qquad \chi_R = \left( \begin{array}{c} 0 \\ v_R \end{array}\right).
\ee
Each one of these vacuum configurations breaks ${\rm SU}(2)_{L,R}\times {\rm U}(1)_{B-L}$ to a ${\rm U}(1)$ subgroup. If $v_R\gg v_L$ the left/right symmetry is spontaneously broken, and we get an explanation of why only the left sector manifests itself at low energies. 

\subsection{A model with only fundamental Higgs fields}

There is also a version of the model with Higgs in the fundamental representations that does not introduce the bi-fundamental Higgs $\Phi$, see e.g. \cite{Hall:2018let}. Indeed, the two VEV's (\ref{fund-H-vev}) are sufficient to break ${\rm SU}(2)_{L}\times{\rm SU}(2)_R\times {\rm U}(1)_{B-L}$ down to the electromagnetic ${\rm U}(1)_{EM}$. The drawback of this model is that it does not allow for Yukawa-type mass terms. However, such terms can be produced using higher dimensional operators with two copies of the Higgs, see \cite{Hall:2018let}. 

\section{Dictionary}
\label{sec:dictionary}

We are now ready to provide an explicit dictionary between elementary particles, organised as in (\ref{fermions}), and elements of $\Oc\times\tilde{\Oc}$.

\subsection{Two commuting complex structures on $\Oc\times \tilde{\Oc}$}

We have already introduced the complex structures $J,\tilde{J}$ on $\Oc,\tilde{\Oc}$ in sections \ref{sec:spin7}, \ref{sec:spin43}. They extend to two commuting complex structures on $\Oc\times \tilde{\Oc}$. 

Let us describe the commutant of $J,\tilde{J}$ acting on $\Oc\times \tilde{\Oc}$ in ${\mathfrak so}(11,3)$. A part of the commutant lives in ${\mathfrak so}(7)\oplus{\mathfrak so}(4,3)$ subalgebra. This is the already described ${\mathfrak so}(6)\subset{\mathfrak so}(7)$ and ${\mathfrak so}(3)\oplus {\mathfrak so}(1,3)\subset{\mathfrak so}(4,3)$. But there are more transformations in the commutant. Indeed, it is easy to see that the generators $E^4\otimes \tilde{E}^i, i=1,2,3$ commute with both $J=E^4$ and $\tilde{J}=-\tilde{E}^1\tilde{E}^2\tilde{E}^3$. These 3 additional generator extend ${\mathfrak so}(3)$ into ${\mathfrak so}(4)$ rotating the directions $e^4, e^{\tilde{1}}, e^{\tilde{2}}, e^{\tilde{3}}$. 

All in all, we see that the commutant of $J,\tilde{J}$ in ${\rm Spin}(11,3)$ is 
\be
G_{J,\tilde{J}} = {\rm SU}(4)\times{\rm SU}(2)_L\times {\rm SU}(2)_R \times {\rm Spin}(1,3),
\ee
i.e., the product of the Pati-Salam and Lorentz groups. We are not careful here about the possible discrete subgroups that are in the kernel of the embedding of this direct product into ${\rm Spin}(11,3)$, as this is of no interest to us in the present paper. 

An even more interesting case is that of the commutant of $J'=R_u$ and $\tilde{J}$. The complex structure $J'$ breaks ${\rm SU}(4)$ to ${\rm SU}(3)\times {\rm U}(1)_{B-L}$. Thus, we get
\be\label{LR-Lorentz}
G_{J',\tilde{J}} = {\rm SU}(3)\times{\rm U}(1)_{B-L}\times {\rm SU}(2)_L\times {\rm SU}(2)_R \times {\rm Spin}(1,3),
\ee
which is the product of the left/right symmetric (\ref{GLR}) and Lorentz groups. 

\subsection{Identification with particles}

The complex structure $J$ splits $\Oc$ into its $(1,0),(0,1)$ subspaces, each of them being a copy of $\C^4$. Moreover, there is a preferred copy of $\C$, arising as the one spanned by $\id, u \in \Oc$. This is where leptons will live. Similarly, the complex structure $\tilde{J}$ splits $\tilde{\Oc}$ into its $(1,0), (0,1)$ subspaces. We know that each of these describes Lorentz 2-component spinors of the same chirality. As in all GUT discussions, we can concentrate on only one of these spaces, thus describing the particle content using Lorentz spinors of the same chirality. 

Putting all of this together, we see that $\Oc\times\tilde{\Oc}$ splits as
\be\label{S-split}
\Oc\times\tilde{\Oc} = L \oplus Q \oplus \bar{L} \oplus \bar{Q}.
\ee
The first two terms here are in the $(1,0)$ subspace of $J$, the second two are in the $(0,1)$ eigenspace. Each of these spaces is also a fundamental representation of ${\rm SU}(2)\subset{\rm Spin}(4,3)$, and a Lorentz spinor. Concentrating on the $(1,0)$ eigenspace of $\tilde{J}$ selects 2-component Lorentz spinors of the same chirality. Thus, the projection of every term in (\ref{S-split}) onto the $(1,0)$ eigenspace of $\tilde{J}$ gives a fundamental representation of ${\rm SU}(2)$ that is also a 2-component Lorentz spinor. 

It is clear that the $Q,\bar{Q}$ factors in (\ref{S-split}) transform as the fundamental and anti-fundamental representation of ${\rm SU}(3)\subset{\rm Spin}(6)\subset{\rm Spin}(7)$, and that all factors in (\ref{S-split}) are charged with respect to the ${\rm U}(1)_{B-L}$, with the assignment of charges being as in (\ref{charges}). It remains to understand the transformation properties of the factors with respect to ${\rm SU}(2)_L\times{\rm SU}(2)_R$. 

The general element of the Lie algebra ${\mathfrak so}(4)$ is given by
\be
 \frac{1}{2} a^i \epsilon^{ijk} \tilde{E}^j\tilde{E}^k + b^i E^4 \otimes \tilde{E}^i.
\ee
This splits into two commuting ${\mathfrak su}(2)$ subalgebras as
\be
\frac{1}{2}(a^i -b^i) \left( \frac{1}{2}  \epsilon^{ijk} \tilde{E}^j\tilde{E}^k - E^4 \otimes \tilde{E}^i  \right) 
+\frac{1}{2}(a^i +b^i) \left( \frac{1}{2}  \epsilon^{ijk} \tilde{E}^j\tilde{E}^k + E^4 \otimes \tilde{E}^i  \right) .
\ee
Let us name the first of these two factors as ${\mathfrak su}(2)_L$, and the second as ${\mathfrak su}(2)_R$.  On the $(1,0)$ eigenspace of $\tilde{J}=-\tilde{E}^1\tilde{E}^2\tilde{E}^3$ we have $\tilde{E}^1\tilde{E}^2=\im \tilde{E}^3$, and so on this subspace the splitting into the two ${\mathfrak su}(2)$ factors becomes
\be
\frac{1}{2}(a^i -b^i) \left( \im   \tilde{E}^i- E^4 \otimes \tilde{E}^i  \right) 
+\frac{1}{2}(a^i +b^i) \left( \im   \tilde{E}^i+ E^4 \otimes \tilde{E}^i  \right) .
\ee
Note that the signs in brackets are reversed onto the $(0,1)$ subspace of $\tilde{J}$. It is now clear that the two eigenspaces of $J=E^4$ transform as $(2,1)$ and $(1,2)$ representations of ${\rm SU}(2)_L\times{\rm SU}(2)_R$. Indeed, on the $(1,0)$ eigenspace of $J$ we can replace $E^4$ by its eigenvalue $-\im$. This makes it clear that this subspace is a singlet with respect to ${\rm SU}(2)_R$. Similarly, the operator $E^4$ has eigenvalue $+\im$ on the $(0,1)$ eigenspace, which makes it clear that it is a singlet with respect to ${\rm SU}(2)_L$. All in all, we get precisely the assignment of representations as in (\ref{charges}). 

It remains to understand the assignment of particles into doublets of the two different ${\rm SU}(2)$. The choice of such assignment for the sector that transforms as the doublet of ${\rm SU}(2)_L$ is a convention. And so we name the components of doublets of ${\rm SU}(2)_L$ as in (\ref{fermions}). These transform as 2-component columns $\chi_L\to U_L \chi_L, U_L\in{\rm SU}(2)_L$. The correct assignment for the right doublets can only be motivated by the Dirac mass terms that couple the left and right sectors. The choices made in (\ref{Phi}), (\ref{Phi-Y}) motivate the assignments as in (\ref{fermions}).

\section{Mass terms}
\label{sec:mass}

We now put the formalism developed to some use. We search for a "Higgs" field that can break the ${\rm Spin}(11,3)$ symmetry, and also be used to write mass terms for the semi-spinor $S_+=\Oc\otimes\tilde{\Oc}$. We thus need a field that lives in the tensor product of the representation $S_+$ with itself. We have the following decomposition of this tensor product 
\be
S_+\times S_+ = \Lambda^1 (\R^{11,3}) \oplus \Lambda^3 (\R^{11,3}) \oplus \Lambda^5 (\R^{11,3}) \oplus \Lambda^7_{sd} (\R^{11,3}).
\ee
Here $\Lambda^k (\R^{11,3})$ is the space of anti--symmetric tensors of rank $k$ in $\R^{11,3}$. The space of 7-forms is not an irreducible representation, but the space $\Lambda^7_{sd}$ of self-dual 7-forms is. The first factor in this sum is anti-symmetric, i.e. projection on it does not vanish only if we assume that the two $S_+$ factors anti-commute. The second term is symmetric. We are interested in the mass terms arising by projecting onto this representation. Thus, we are interested in the symmetric bilinear forms on $S_+$ that can be written as
\be
C^{\mu\nu\rho} \lla \Psi, \Gamma_\mu \Gamma_\nu \Gamma_\rho\Psi \rra.
\ee

\subsection{Projection onto $\Lambda^3 \R^{11,3}$}

Three copies of $\Gamma$-matrices can be inserted between two $S_+$ states and produce a ${\mathfrak so}(11,3)$ invariant expression. The different arising components are 
\be
\lla \Psi, \Gamma_{x_1 x_2 x_3} \Psi \rra = \lla \Psi, L_{x_1} L_{x_2} L_{x_3} \otimes \id \Psi \rra, 
\quad
\lla \Psi, \Gamma_{y_1 y_2 y_3} \Psi \rra = - \lla \Psi, \id \otimes L_{y_1} L_{y_2} L_{y_3}  \Psi \rra, 
\\ \nonumber
\lla \Psi, \Gamma_{x_1 x_2 y_1} \Psi \rra = \lla \Psi, L_{x_1} L_{x_2}  \otimes  L_{y_1} \Psi \rra, 
\quad
\lla \Psi, \Gamma_{x_1 y_1 y_2} \Psi \rra = - \lla \Psi,   L_{x_1} \otimes L_{y_1} L_{y_2}  \Psi \rra.
\ee
This means we have the following expression for the projection $S_+\times S_+ \to \Lambda^3 \R^{11,3}$
\be\label{364-rep}
\lla \Psi, \Gamma\Gamma\Gamma \Psi \rra = \lla \Psi, E^i E^j E^k \otimes \id \Psi \rra e^{ijk} + \lla \Psi, E^i E^j \otimes  \tilde{E}^k  \Psi \rra e^{ij} \tilde{e}^k \\ \nonumber
-\lla \Psi,   E^i \otimes \tilde{E}^j \tilde{E}^k  \Psi \rra e^i \tilde{e}^j \tilde{e}^k 
-\lla \Psi, \id \otimes \tilde{E}^i \tilde{E}^j \tilde{E}^k   \Psi \rra \tilde{e}^i \tilde{e}^j \tilde{e}^k.
\ee

\subsection{ ${\rm SU}(3)\times{\rm U}(1)$ and Lorentz invariant pairing}

We now select among the 364 terms in (\ref{364-rep}) those that are ${\rm SU}(3)$, ${\rm U}(1)_{EM}$ and Lorentz invariant. 

We have the ${\rm SU}(3)$ invariant 3-forms in $\R^6\subset{\rm Im}(\Oc)$. These arise as the real and imaginary parts of an ${\rm SU}(3)$ invariant $(3,0)$ form
\be
\Omega := ( e^1 - \im e^5)(e^2-\im e^6) (e^3-\im e^7).
\ee
Note, however, that $\Omega$ is a singlet with respect to ${\rm SU}(2)_L\times{\rm SU}(2)_R$, but transforms non-trivially under the ${\rm U}(1)_{B-L}$. This means that $\Omega$ is not a singlet with respect to ${\rm U}(1)_{EM}$, and cannot be used. There is then an ${\rm SU}(3)\times{\rm U}(1)_{B-L}$ invariant real 2-form
\be
\omega = e^{15} + e^{26}+e^{37}.
\ee
The last building block we are allowed to use in the $\R^7$ factor is the 1-form $e^4$. 

For objects in ${\rm Im}(\tilde{\Oc})$, we are only allowed to use the 1-, 2- and 3-forms built from $\tilde{e}^{1,2,3}$. These 1- and 2-forms transform as vectors under the weak ${\rm SU}(2)$, while the 3-form is invariant. We can use objects that transform non-trivially under the ${\rm SU}(2)$ because we do not intend to preserve this symmetry. In other words, objects that transform non-trivially under the ${\rm SU}(2)$ are components of a "Higgs" field that we know is present in the mass terms of the SM.

This leads us to consider the following "mass" quadratic form in $S_+$
\be\label{mass}
\lla \Psi, \Psi\rra_M = m_1 \lla \Psi, ( E^1 E^5 + E^2 E^6+E^3 E^7) E^4 \otimes \id \Psi \rra 
- m_2 \lla \Psi, \id \otimes \tilde{E}^1 \tilde{E}^2 \tilde{E}^3 \Psi \rra \\ \nonumber
+ \psi^i \lla \Psi , ( E^1 E^5 + E^2 E^6+E^3 E^7) \otimes \tilde{E}^i \Psi \rra
+ \frac{1}{2} \phi^i \epsilon^{ijk} \lla \Psi, E^4 \otimes \tilde{E}^j \tilde{E}^k \Psi\rra.
\ee
Here the indices $i,j,k$ run over the values $1,2,3$, and summation convention is implied. Here $\xi\in\C$ is a complex parameter, $m_{1,2}\in\R$ and $\phi^i,\psi^i$ are two real vectors.

\subsection{Evaluating the mass terms}

We now  evaluate (\ref{mass}) on the particle states. Every element of $\Oc$ splits into 8 different states $L,\bar{L}, Q_i,\bar{Q}_i$, and we have computed the relevant pairings in this basis in (\ref{O-formulas-1}), (\ref{O-formulas-2}), (\ref{Ot-formulas-1}), (\ref{Ot-formulas-2}). 

Let us concentrate on the "quark" sector first. We get
\be
\left( \begin{array}{cc} u & d \end{array}\right)  \left( \begin{array}{cc} 0 & 1 \\ -1 & 0 \end{array}\right)  \Phi_Q \left( \begin{array}{c} \bar{d} \\ -\bar{u} \end{array}\right) = Q^T \epsilon \Phi_Q \bar{Q} ,
\ee
where
\be
\Phi_Q = \left((m_1+\im m_2)\id + (\phi^i -\im \psi^i) \sigma^i \right) ,
\ee
and $\sigma^i$ are the usual Pauli matrices. This should be compared with (the quark version of) (\ref{Phi-Y}).

For the lepton sector we get similarly
\be
\left( \begin{array}{cc} \nu & e \end{array}\right)  \left( \begin{array}{cc} 0 & 1 \\ -1 & 0 \end{array}\right)  \Phi_L \left( \begin{array}{c} \bar{e} \\ -\bar{\nu} \end{array}\right) = L^T \epsilon \Phi_L \bar{L}  ,
\ee
where
\be
\Phi_L =  \left((-3m_1+\im m_2)\id + (\phi^i +3\im \psi^i) \sigma^i \right) .
\ee
It is clear that we get a relation between the masses of quarks and leptons, as is typical in GUT, but the particular relation that arises does not seem to be of any phenomenological interest. 

The main conclusion of the calculation performed is that the bi-doublet Higgs field $\Phi$ of the left/right symmetric version of the SM that appears in the Yukawa mass terms of this model (and thus contains the familiar Higgs field) can be identified with a general ${\rm SU}(3)\times{\rm U}(1)_{EM}$ and Lorentz invariant 3-form in $\R^{11,3}$. It is interesting that this 3-form Higgs field gives masses to both leptons and quarks, and breaks the symmetry between them. Note, however, that while the 3-form (\ref{364-rep}), with values of the parameters chosen to align with (\ref{Phi-vev}), breaks ${\rm Spin}(11,3)$ to the product of ${\rm SU}(3)\times{\rm U}(1)_{B-L}$ and the Lorentz groups, there is also the diagonal ${\rm U}(1)\subset{\rm SU}(2)_L\times{\rm SU}(2)_R$ that survives, similar to the left/right symmetric model. Thus, the 3-form Higgs field by itself is not sufficient to break ${\rm Spin}(11,3)$ to the product of the unbroken part of the SM gauge and Lorentz groups, as ${\rm U}(1)_{B-L}$ survives. As in the left/right symmetric model, other Higgs fields are required to complete the breaking.

\section{Discussion}

We have developed a very efficient formalism that describes the semi-spinor representation of ${\rm Spin}(11,3)$ as the direct product of the usual $\Oc$ and split $\tilde{\Oc}$ octonions. This is likely the most efficient description of this group, as is in particular manifested by the very simple description (\ref{lie-11-3}) of the Lie algebra ${\mathfrak so}(11,3)$. This formalism, together with its other applications covered by the tables in (\ref{table}), likely has applications other from the considered here particle physics context.

The main application of our representation theoretic construction is based on the fact that components of the fermionic fields of one generation of the SM can be identified with the components of the semi-spinor representation $S_+$ of ${\rm Spin}(11,3)$. We then introduced two commuting complex structures $J,\tilde{J}$ on $S_+$ whose commutant is the product (\ref{LR-Lorentz}) of the left/right symmetric and Lorentz groups. The fact that there is naturally two commuting complex structures is simply explained by the product structure $S_+=\Oc\otimes\tilde{\Oc}$, because $J,\tilde{J}$ arise as complex structures on $\Oc,\tilde{\Oc}$ respectively. The space $S_+$ then splits into its $(1,0),(0,1)$ eigenspaces with respect to both of the complex structures, and we saw that this splitting for $J$ corresponds to the splitting of fermionic states into particles and anti-particles, denoted by unbarred and barred letters in (\ref{S-split}). The splitting for $\tilde{J}$, on the other hand, corresponds to the splitting of Lorentz 4-component spinors into two different types of 2-component spinors, or into two different chiralities. What arises is the structure of the SM states as described by the left/right symmetric version of the SM, see (\ref{fermions}). Thus, the developed here octonionic model achieves the ultimate "kinematic" unification of all the fermionic states, including the Lorentz spin states. It is hard to imagine a description of all these states more elegant than the one provided by our octonionic model. 

Nevertheless, as soon as one brings into play the group as large as ${\rm Spin}(11,3)$, we are faced with the all important question of what breaks this large symmetry to the ones we see in Nature. The provided description of the needed breaking pattern that uses the two complex structures $J,\tilde{J}$ is elegant, but it brings us no closer to answering the question about a mechanism for this symmetry breaking. The situation is similar in GUT theories, where one introduces a large symmetry group such as ${\rm Spin}(10)$, and then needs to provide a mechanism for breaking this symmetry to the groups we observe in Nature. However, in the GUT schemes we have the Higgs mechanism, which we are confident is the correct theoretical description of the symmetry breaking at least in the case of the electroweak theory. In this sense, the case of ${\rm Spin}(11,3)$ is very different because we can no longer resort for help to the usual Higgs mechanism with some collection of Higgs fields. Indeed, it is clear that we have put together what all textbooks say one should never mix - the group of gauge transformations acting on the "internal" particle degrees of freedom, and the Lorentz group that acts by spacetime rotations. There is certainly no established mechanism that can break the group mixing such different symmetries together. Some attempts at such a mechanism are known under the name of "spontaneous soldering", see e.g. \cite{Percacci:1984ai}, \cite{Nesti:2009kk}, but their status is very different from the established Higgs mechanism of particle physics. 

The only thing we can say in the defence of the whole approach is that the similarity in the two complex structure $J,\tilde{J}$, the first of which is needed to go from ${\rm Spin}(10)$ to the Pati-Salam group, and the second needed to see the Lorentz group arising, is so striking that this suggests that the mechanism selecting them must be one and the same. In the case of ${\rm Spin}(10)$ to the Pati-Salam breaking one can resort to the usual Higgs mechanism, with the representation ${\bf 210}$ needed for this purpose. However, one then needs other ${\rm Spin}(10)$ representations to break the symmetry even further. There are many possible arising models, and these ambiguities is one of the biggest weaknesses of the ${\rm Spin}(10)$ unification. It may then seem that we have enlarged the group even further, and so have even more possibilities for the symmetry breaking. But this is unlikely the case, because we don't know of {\bf any} mechanism that would select the Lorentz group producing the complex structure $\tilde{J}$. It is then possible that the necessity of having a mechanism that produces $J,\tilde{J}$ at the same time is the much desired constraint that limits ambiguities also of the ${\rm Spin}(10)$ unification. All this, however, remains a speculation, as there is no required mechanism as of yet. 

As a small step towards the mechanism required, we described the symmetry breaking using the 3-form representation $\Lambda^3(\R^{11,3})$. This arises as the representation of the smallest dimension in the second symmetric power of the semi-spinor representation with itself. We have seen that a particular 3-form (\ref{364-rep}) is a singlet with respect to ${\rm SU}(3)\times{\rm U}(1)_{B-L}$ and Lorentz groups, and transforms as the bi-doublet representation of ${\rm SU}(2)_L\times{\rm SU}(2)_R$. Selecting a VEV for this field as in the left/right symmetric version of the SM, see (\ref{Phi-vev}), we can break ${\rm Spin}(11,3)$ to the product of ${\rm SU}(3)\times{\rm U}(1)_{B-L}$, Lorentz, and the diagonal ${\rm U}(1)\subset {\rm SU}(2)_L\times{\rm SU}(2)_R$. But this "Higgs" field, although a step in the right direction, does not break the remaining ${\rm U}(1)_{B-L}\times{\rm U}(1)$ down to the electromagnetic ${\rm U}(1)_{EM}$. Nevertheless, this construction does suggest that a field in $\Lambda^3(\R^{11,3})$ is the most natural one to attempt to produce a dynamical symmetry breaking mechanism. What is far from clear, however, if there is dynamics that can select precisely the configuration (\ref{364-rep}) for this field. In particular, it is very important that no Lorentz symmetry breaking terms get produced, and it is far from certain that this can be achieved. More work is needed to explore all these ideas. 

Our final comment is about the ${\rm Spin}(7,7)$ model advocated in our previous work \cite{Krasnov:2018txw}. This work used the formalism of polyforms to describes spinors. The main result was the observation that the Weyl-Dirac kinetic term 
\be\label{Dirac}
\int \lla \Psi, D \Psi \rra,
\ee
where $D$ is the chiral Dirac operator that maps $S_+\to S_-$ and $\Psi\in S_+$, when dimensionally reduced to $1+3$ dimensions reproduces the collection of correct Dirac kinetic terms for all the SM fermions. The calculation was based on the embedding ${\rm Spin}(6)\times{\rm Spin}(1,3)\times{\rm Spin}(4)\subset{\rm Spin}(7,7)$. Our first remark is that this observation applies also to the considered here case of ${\rm Spin}(11,3)$. Thus, also in the ${\rm Spin}(11,3)$ case the dimensional reduction of (\ref{Dirac}) from $\R^{11,3}$ to $\R^{1,3}$ reproduces the kinetic terms for all the SM fermions, thus "unifying" them in the single term (\ref{Dirac}). Our second remark is that the described here octonionic formalism for both ${\rm Spin}(7,7)$ and ${\rm Spin}(11,3)$ makes it clear that the later case is far superior to the former for the purpose of describing the elementary particles. Indeed, it is clear that the Pati-Salam and left/right symmetric groups together with Lorentz fit into ${\rm Spin}(11,3)$ much more naturally then into ${\rm Spin}(7,7)$. As we saw in this paper, the factor $\Oc$ naturally gives rise to the strong ${\rm SU}(3)$ and ${\rm U}(1)_{B-L}$ groups, while the $\tilde{\Oc}$ factor gives the "weak" ${\rm SU}(2)$ and Lorentz groups. There is no such simple description in the split case ${\rm Spin}(7,7)$, and so it must be concluded that from the two pseudo-orthogonal groups in 14 dimensions that have real semi-spinor representations, it is the ${\rm Spin}(11,3)$ case that is relevant for the purpose of describing the elementary particles, not ${\rm Spin}(7,7)$.

\end{document}